\documentstyle[12pt,graphics]{article}

\newfont{\headfont}{cmbx10 scaled 1440}
\newfont{\headfontb}{cmbx10 scaled 1200}
\newfont{\namefont}{cmr9}
\newfont{\initialfont}{cmr9 scaled 1200}
\newfont{\addfont}{cmti10}

\begin{document}

\begin{titlepage}
\renewcommand{\thefootnote}{\fnsymbol{footnote}}
\begin{center}
{\headfont  Decoherence of Black Holes by Hawking Radiation}
\end{center}
\vskip 0.3truein
\begin{center}
{ {\initialfont J}{\namefont EAN-}{\initialfont G}{\namefont UY}
                    {\initialfont D}{\namefont EMERS  } }
\end{center}
\begin{center}
{\addfont{ Department  of Physics, McGill University}}\\
{\addfont{ Ernest Rutherford  Building, 3600 University Street}}\\
{\addfont{ Montr\'{e}al, PQ, Canada H3A 2T8}}\\
\end{center}

\vspace{3mm}

\begin{center}
{ {\initialfont C}{\namefont LAUS}
                    {\initialfont K}{\namefont IEFER  } }
\end{center}
\begin{center}
{\addfont{ Fakult\"at f\"ur Physik, Universit\"at Freiburg}}\\
{\addfont{ Hermann - Herder - Strasse 3}}\\
{\addfont{ D-79104 Freiburg, Germany}}\\

\end{center}

\vskip 0.5truein
\begin{abstract}
We discuss in detail the semiclassical approximation for the CGHS model
of two-dimensional dilatonic black holes. This is achieved by a formal
expansion of the full Wheeler-DeWitt equation and the momentum
constraint in powers of the gravitational constant. In highest order,
the classical CGHS solution is recovered. The next order yields a
functional Schr\"odinger equation for quantum fields propagating on this
background. We show explicitly how the Hawking radiation is recovered
from this equation. Although described by a pure quantum state,
the expectation value of the number operator exhibits a Planckian
distribution with respect to the Hawking temperature.
 We then show how this  Hawking
radiation can lead to the decoherence of black hole superpositions.
The cases of a superposition of a black hole with a white hole,
as well as of a black hole with no hole, are treated explicitly.
\end{abstract}
\vskip .4truein
\vfill
\leftline {McGill 95-56 }
\leftline {Freiburg THEP-95/22}
\smallskip
\end{titlepage}
\vskip .5in

\section{Introduction}

One of the main applications of any quantum theory of gravity
should be the complete understanding of black hole evaporation.
Since such a theory is not yet available, attention has been focused
on simpler models whose quantization is expected to be tractable.
Most notably among these are the models of two-dimensional
dilaton gravity (see, for example, \cite{stro} for a detailed
review). Although much insight has been gained into the evaporation
process and, in particular, the back reaction of the Hawking radiation
onto the black hole and the problem of ``information loss",
a full understanding of the complete, non-perturbative, evolution
remains elusive. The reason for this is that even for such models
the full quantum theory is not known.

At present, the most popular approaches towards the quantization of
gravity are superstrings and quantum general relativity.
While many of the models discussed are in fact ``string-inspired",
a non-perturbative understanding of the full string field theory
is not in sight. Effort is therefore also concentrated on a more
conservative approach: The application of canonical quantization rules
to the general theory of relativity. While this fails to yield,
in four spacetime dimensions, a viable theory in a perturbative sense,
it might nevertheless make sense non-perturbatively. Even if this
were not the case, this framework appears appropriate to exhibit
the main physical problems and to suggest methods for their solution.
We shall thus restrict our investigation to this framework, too.

What progress has been made towards a quantum theory of black holes
within canonical quantum gravity? The situation of a spherically
symmetric black hole without matter has been completely solved
\cite{K,KT}. After the embedding variables describing the location
of three-dimensional hypersurfaces in spacetime are isolated,
the remaining state is a quantum mechanical wave function,
$\psi(M)$, where $M$ is the mass of the hole. The variable conjugate
to $M$ is related to the time in the asymptotic region. The fact that
no field theoretic degrees of freedom are present reflects of course
the situation of the classical theory (Birkhoff's theorem).
In more complicated models where matter degrees of freedom are present
it seems no longer possible to proceed as in \cite{K,KT} and to isolate the
embedding variables from the ``true" degrees of freedom \cite{rom}.
We shall thus stick to the approach where all degrees of freedom
are treated on an equal footing as dynamical variables.

Instead of finding the full solution in simple situations,
approximation schemes have been employed in the full theory.
One example is a formal expansion of the full wave functional
in powers of the gravitational constant $G$ \cite{kiefer1}.
The $G$-expansion scheme allows one to derive the
limit of quantum field theory in a background spacetime as well as
to find quantum gravitational correction terms to this limit.

One application of such correction terms for black holes was
made in \cite{KMS} where it was shown that some of these terms
yield non-unitary contributions which become relevant if the
black hole mass approaches the Planck mass. This could be of central
importance for the information loss problem. The investigation in
\cite{KMS} was, however, only heuristic, since it did not pay
attention to problems of regularization.
One of the motivations for the present paper is the attempt to
study the semiclassical expansion in the much simpler context of
two-dimensional dilatonic gravity. This will enable us, in particular,
to understand the emergence of Hawking radiation in this
framework, which thus complements the standard derivation with the use
of Bogolubov coefficients.

Furthermore, it will also be straightforward to study the validity of
the limit where spacetime is semiclassical and matter fully quantum.
This question received some interest recently because it was found
in \cite{klmo} that semiclassical gravity breaks down on hypersurfaces
which capture both the infalling matter near the horizon and
the Hawking radiation. In our paper we shall consider superpositions
of different semiclassical components and investigate to which extent
they become dynamically independent. The key ingredient is thereby
played by the notion of decoherence - the emergence of quasiclassical
properties through the irreversible interaction with the
environment (see, for example, \cite{kiefer1} and the references
therein). What can possibly play the role of the environment for the
black hole? It is the Hawking radiation itself, which
may serve as the decohering agent. Although it may be weak,
it possesses a large entropy capacity due to its many degrees of freedom
and can thus carry all the information about the superpositions
which thereby decohere.

Our paper is organized as follows.
In Section~2 we shall review the canonical formalism for
the CGHS model of dilatonic gravity, in particular the corresponding
Wheeler-DeWitt equation and the quantum momentum constraint.
Section~3 is then devoted to a detailed discussion of the semiclassical
approximation. The highest order yields the Hamilton-Jacobi equation
for dilatonic gravity with\-out matter fields.
 We show how the classical CGHS solution
can be recovered from a solution to this equation. We also comment on
a connection between this solution and the entropy of the black hole.
We then derive in Section~4
 the functional Schr\"odinger equation for matter fields
on the semiclassical background. An important part of our paper
is then devoted to a derivation of Hawking
radiation from this Schr\"odinger equation. We consider both the case
of a collapsing geometry and of an eternal black hole.
We shall find that Hawking radiation is obtained from a pure
quantum state in the sense that the expectation value of the number
operator for field quanta exhibits a thermal spectrum with respect
to the Hawking temperature. We end this section with a brief discussion
of quantum gravitational correction terms to the functional
Schr\"odinger equation.
 In Section~5 we discuss in detail
how Hawking radiation can lead to decoherence for superpositions
of black hole states. In particular, we study the superposition
of a black hole with a white hole as well as the superposition of
a black hole with no hole. The results then justify the separate
consideration of each semiclassical component, since they dynamically
decouple from each other. Finally, Section~6 contains a brief
conclusion as well as an outlook on future work.

\section{Quantum theory of two-dimensional black ho\-les}
 \label{Unruh}

In this section we shall apply canonical quantum gravity to
spherically symmetric black holes. From the technical point of view,
this lies between the full quantum theory and finite-dimensional
minisuperspace models, since the wave functional depends now on
one-dimensional {\em fields}. An effective two-dimensional theory
can thus be obtained by imposing a spherical symmetric ansatz
for the classical four-dimensional metric and applying the
standard ADM procedure for the remaining degrees of freedom.

Recently, however, a slightly modified two-dimensional theory has
received considerable attention \cite{cghs}. Apart from being
``string inspired", the reason for the choice of
 this ``CGHS model" is basically simplicity,
giving rise to the hope that the full quantum evolution (including
the back reaction of Hawking radiation on the gravitational
background) can be tackled in this framework. For this reason we
restrict our attention to this model, too, but keep in mind that
this will give at best some hints of how to proceed in the full
theory.

The action for the two-dimensional gravity model of CGHS coupled to
$N$ massless scalar fields reads
\begin{equation}
S=\frac{1}{G}\int dxdt\:\sqrt{-\bar{g}} \:e^{-2\bar{\phi}}
(\bar{R}+4(\bar{\nabla}
 \bar{\phi})^2+
4\lambda^2)-\frac{1}{2}\int dxdt\:\sqrt{-\bar{g}}\:
 \sum_{i=1}^N(\bar{\nabla}f_i)^2,
\label{cghs}
\end{equation}
where $\bar{\phi}$ is the dilaton field, $f_i$ are the matter fields and
$G$ is the (unitless) gravitational coupling constant, which
we have introduced for later convenience. We note that in the
corresponding model resulting from dimensional reduction the term
$4\lambda^2$ is replaced by $e^{2\bar{\phi}}/l^2$, where $l^2$ is a relic
of the curvature components orthogonal to the two-sphere, which
usually is measured in terms of the Planck length.
In this sense the two-dimensional ``cosmological constant term"
is an artefact which has its origin in the ``real" (four-dimensional)
gravitational constant. Alternatively, $\lambda^{-1}$ may be viewed as being
related to the magnetic charge of four-dimensional dilatonic
black holes.

In order to study the canonical structure of (\ref{cghs}), we make the
transformation $\phi=e^{-2\bar{\phi}}$ and
 $g_{\alpha\beta}=e^{-2\bar{\phi}} \bar{g}_{\alpha\beta}$, which eliminates the
kinetic
term for the dilaton. This yields
\begin{equation}
S=\int dxdt\:\sqrt{- {g}} \: \left[\frac{1}{G}(R\phi+4\lambda^2)-\frac{1}{2}
(\nabla f)^2 \right],
\label{sk}
\end{equation}
where we have written out only one of the scalar fields for simplicity.
Following \cite{kuns1} we write the metric in an ADM-like parametrization:
\begin{equation}
ds^2=e^{2\rho}\: [-\sigma ^2dt^2+(dx+\xi dt)^2]\ , \label{met}
\end{equation}
where $\sigma $ denotes the lapse and $\xi$ the shift function
(note the additional
factor $e^{2\rho}$ in the definition of the metric, which differs from
the standard ADM convention).
The action then reads, upon neglecting surface terms\footnote{The
surface term describing the ADM energy,
which is required in the Hamiltonian in order
to consistently recover Hamilton's equations of motion, is not needed
for the constraint analysis below and can thus consistently be
neglected at this stage.}

\begin{equation}
S= \int dxdt\:( \dot{\phi} \Pi_\phi+\dot{\rho} \Pi_\rho+\dot{f} \Pi_f -
\xi{\cal F}-\sigma  {\cal G}),
\end{equation}
where
\begin{eqnarray}
\Pi_\phi&=&\frac{2}{\sigma  G}(\xi\rho^\prime+\xi^\prime-\dot{\rho}),
\nonumber \\ \relax
\Pi_\rho&=&\frac{2}{\sigma  G}(\xi\phi^\prime -\dot{\phi}), \nonumber \\ \relax
\Pi_\sigma & \approx &0, \; \; \; \Pi_{\xi}\approx 0, \nonumber \\ \relax
\Pi_f&=&\frac{1}{\sigma }(\dot{f} - \xi \phi^\prime)\ .
\label{mom}
\end{eqnarray}
Here,  primes and dots respectively  denote  a derivative with respect
to $x$ and $t$, while ``$\approx 0 $" is used  to  specify the location
of the constraint surface in  phase space.
The momentum and Hamilton ian constraints are then given by the expressions
\begin{eqnarray}
{\cal F}&=&\rho^\prime \Pi_\rho -\Pi_\rho^\prime +   \phi^\prime \Pi_\phi +
\phi^\prime \Pi_ f
    \approx 0, \nonumber \\ \relax
{\cal G}&=&\frac{2}{G} V_G-\frac{G}{2}
\Pi_\phi\Pi_\rho+\frac{1}{2}  \Pi_ f^2+V_M \approx 0 \ .
\label{cons}
\end{eqnarray}
where
\begin{equation}V_G=4(\phi^{\prime\prime} -\phi^\prime\rho^\prime -
2\lambda^2e^{2\rho}),\;
    V_M=\frac{1}{2}f^{\prime 2}\ . \end{equation}
The fields $\rho$ and $\phi$ don't enter the momentum constraint
on the same footing, since the former transforms as a density,
whereas the latter transforms as a scalar.
Quantization then proceeds \`a la Dirac by imposing these constraints
as restrictions on physically allowed wave functionals
$\Psi[\rho(x),\phi(x),f(x)]$ and assuming  the usual equal-time commutation
relations, $[\rho(x),\Pi_{\rho}(y)]=i\delta(x-y)$ etc., are obeyed.
Note that this may
not be consistent, since $\phi$ is restricted, by the above
redefinition, to positive values. Affine commutation relations would
thus be more appropriate. However, we do not
expect this difference to be relevant for the present discussion
which focuses on a semiclassical expansion. Note also that
this quantization procedure differs from methods recently employed
in this context \cite{K}, where the constraints are first manipulated
on the classical level to isolate explicitly the embedding
variables (which have to be spacetime scalars), with respect to which
a Schr\"odinger equation is then obtained after quantization.

 Since the supermetric is {\em flat}
in these field variables, there is no factor ordering ambiguity
in the Hamiltonian constraint, and one readily obtains
\begin{eqnarray}
{\cal H}_{\parallel}\Psi & = & \left(\rho^\prime\frac{\delta}{\delta\rho}
 -\frac{d}{dx}\frac{\delta}{\delta\rho} +\phi^\prime\frac{\delta}{\delta\phi}
 +f^\prime\frac{\delta}{\delta f}\right)\Psi=0, \label{mc} \\
{\cal H}_{\perp}\Psi & = & \left(\frac{G}{2}
\frac{\delta^2}{\delta\rho\delta\phi}
 -\frac{1}{2}\frac{\delta^2}{\delta f^2} +\frac{1}{2G}V_G +V_M\right)
 \Psi =0\ . \label{wdw}
\end{eqnarray}

On a first glance, (\ref{wdw}) seems to describe an almost trivial field
theory, the only ``interaction" coming from the Liouville term
proportional to $e^{2\rho}$. In particular, the field $f$ has totally
decoupled from the gravitational fields $\rho$ and $\phi$, a consequence
of the conformal coupling in two dimensions. However, the presence of
the momentum constraints induces a ``correlation interaction"
\cite{CJZ} which prevents the formulation of solutions separating
in the respective variables. A major problem also seems to be
the occurrence of anomalies which spoil the validity of
(\ref{mc}) and (\ref{wdw}) unless a quantum modification of the
Hamiltonian is made \cite{CJZ}. Since this may be of minor
relevance for the topic of this paper, these problems will not
be discussed here. We hope, however, to return to these issues
in a future publication.

\section{The semiclassical expansion} \label{hja} 

Since it is not yet clear how the full equations (\ref{mc}) and
(\ref{wdw}) can be addressed rigorously, we turn here to
a semiclassical expansion scheme.
 Such a scheme is also interesting on its own,
as Hawking radiation can be properly understood in that context.
This seems indeed appropriate
as long as one focuses on the evolution of quantum fields
on a semiclassical gravitational background \cite{kiefer1}.
More precisely, this expansion is of the Born-Oppenheimer type
with respect to the gravitational constant. It is for this purpose
that we have introduced the dimensionless constant $G$ in
(\ref{cghs}).
Alternatively, one can perform this expansion with respect to
the -- large -- number $N$ of matter fields,
where one starts from a Liouville-type action in which a nonlocal
redefinition of the fields in (\ref{cghs} has been preformed \cite{AI}.
 The appropriate parameter there turns out to be
$\kappa\equiv (N-24)/6$.
Since in \cite{AI}, $\kappa$ takes the same place in the Hamiltonian
constraint that  in our case is taken by $G$,
both schemes are formally equivalent.
We choose to perform an expansion with respect to $G$.
We note, however, that this scheme is different from the one
where the semiclassical Einstein equations (with the expectation
value of the energy-momentum tensor on the right-hand side) are
discussed in the limit of large $N$ \cite{stro}. While this can
serve to suppress graviton loops compared to matter loops, the
present approximation scheme takes also into account quantum
gravitational correction terms, see Section~4.4.

We thus assume an ansatz of the form
\begin{equation}
\Psi[\rho,\phi,f]=e^{i({G}^{-1}S_0+S_1+GS_2+ \ldots)}.
\end{equation}
At order $G^{-2}$ one finds ${\delta S_0}/{\delta f}=0$, while at order
$G^{-1}$
we obtain:
\begin{equation}
\frac{\delta S_0}{\delta \phi} \frac{\delta S_0}{\delta \rho} =
 4(\phi^{\prime\prime}-\phi^\prime \rho^\prime-2\lambda^2 e^{2\rho}) =V_G,
\label{hje}
\end{equation}
which is the Hamilton- Jacobi equation for pure gravity.
The solution of (\ref{hje}) was found in \cite{kuns1} to be
\begin{equation}
S_0=\int dx\: \left [Q+\phi^\prime \ln\left(
\frac{2\phi^\prime-Q}{2\phi^\prime+Q}
 \right)\right] \label{s0}
\end{equation}
with $Q=2\sqrt{{\phi^\prime}^2 +(C-4\lambda^2\phi ) \:e^{2\rho}}$,
 where $C$ is an integration
constant. In fact, $C$ is a constant for the full set of constraints
in the {\em pure} gravity case, since it can be shown \cite{kuns1}
that the spatial derivative of the functional
\begin{equation} C[\rho,\Pi_{\rho},\phi]\equiv
e^{-2\rho}\left(\frac{\Pi_{\rho}^2}
 {4}-\phi^{\prime 2}\right) +4\lambda^2\phi \label {ob} \end{equation}
 is equal to a linear combination of the constraints.
Note that Eq.(\ref{s0}) obeys ${\delta S_0}/{\delta \phi}={V_G}/{Q}$ and
 $ {\delta S_0}/{\delta \rho}= Q $.
Moreover, $C$ commutes with all constraints and can thus be interpreted
as an ``observable" -- in fact, it is proportional to the
ADM mass (see below). An analogous quantity appears in the reduced
models from four dimensions \cite{K,KT}. In the presence of matter
fields, the spatial derivative of the functional $C$ no longer vanishes
on the constraint surface. Therefore, the general approach to reduction
made in \cite{K,KT} is no longer applicable here, see \cite{rom}.

It is also easily checked that $S_0$ obeys the momentum constraint
in this order of approximation (since this constraint  does not
contain $G$, its expansion is straightforward).
 Note that $S_0$ becomes imaginary if either $Q^2<0$ or
 $4{\phi^\prime}^2-Q^2<0 \Leftrightarrow C>4\lambda^2\phi$ (both conditions
cannot be satisfied simultaneously). For the black hole solution
discussed below, the latter condition describes the region
inside the horizon.

\subsection{Spacetime and dilaton}
We now recover explicitly the classical solutions for the
 conformal factor of the metric as well as for the dilaton,
 $\rho(x,t)$ and  $\phi(x,t)$, from the solution (\ref{s0})
 to the Hamilton-Jacobi equation.
 The momenta are given in (\ref{mom}), and using
$\Pi_\theta={G}^{-1}{\delta S_0}/{\delta \theta}$ for $\theta=\rho$ and
$\phi$ one has
\begin{eqnarray}
\frac{2}{\sigma}(\xi \rho ^\prime+\xi^ \prime -\dot{\rho})&=
&\frac{\delta S_0}{\delta \phi}=
\frac{V_G}{Q}\ , \label{fd} \\
\frac{2}{\sigma}(\xi \phi^ \prime -\dot{\phi})&=&\frac{\delta S_0}
{\delta \rho}=Q\ .
\label{rd}
\end{eqnarray}
It turns out to be convenient to work in the
 the conformal gauge, i.e. we set $\sigma =1$ and $\xi=0$.
Using in addition
lightcone variables with $x^\pm=t\pm x$ and squaring (\ref{fd})
yields
\begin{equation}
4\partial_-\phi\: \partial_+\phi= (C-4\lambda^2 \phi)e^{2\rho} \ .
\label{1cond}
\end{equation}
Now using (\ref{fd}) in
(\ref{rd}) gives $\dot{\rho}\:\dot{\phi}={V_G}/{4}$
leading to
\begin{equation}
(\partial_+^2\phi -2 \partial_+\phi \: \partial_+\rho)+(\partial_-^2\phi -
2 \partial_-\phi \: \partial_-\rho )=
2(\lambda^2 e^{2\rho}+\partial_-\partial_+\phi) \label{cat}  \ ,
\end{equation}
 while $\partial_-$ and $\partial_+$ of Eq.  (\ref{1cond}) gives
\begin{eqnarray}
-\partial_-\phi \: (\lambda^2 e^{2\rho}+\partial_-\partial_+\phi)&=&
\partial_+\phi \:(\partial_-^2\phi -2 \partial_-\phi\: \partial_-\rho )
\label{cinq} \ ,\\
-\partial_+\phi \: (\lambda^2 e^{2\rho}+\partial_-\partial_+\phi)&=&
\partial_-\phi \:(\partial_+^2\phi -2 \partial_+\phi \: \partial_+\rho )
\label{six} \ .
\end{eqnarray}
{}There are a priori two ways to solve (\ref{cat}), (\ref{cinq}),
and (\ref{six}).
The first is to set $\partial_+\rho=-\partial_-\rho$ and  $\partial_+\phi=-
\partial_-\phi$,
corresponding
to static solutions $\dot{\phi}=\dot{\rho}=0$. The second is to assume
that the quantities
in each of the parenthesis vanishes. This is what we now do;  it
will become clear that this assumption does not exclude the
static solutions.
 Thus,
\begin{eqnarray}
 e^{2\rho}\partial_-(e^{-2\rho}\partial_-\phi)&=&0 \ ,\nonumber \\ \relax
 e^{2\rho}\partial_+(e^{-2\rho}\partial_+\phi)&=&0
\label{29}
\end{eqnarray}
or
\begin{eqnarray}
 \partial_-\phi &=&  e^{2\rho} A(x^+) \ ,\nonumber \\ \relax
 \partial_+\phi &=&  e^{2\rho} B(x^-) \ .
\label{30}
\end{eqnarray}
Here,  the functions $A(x^+)$ and $B(x^-)$ are
non-vanishing and  have opposite signs, but are  otherwise
arbitrary.
  Making use of (\ref{30}), the solution for $\phi$  in (\ref{1cond})
may be written
\begin{equation}
\frac{C}{4}-\lambda^2 \phi= u \ e^{-\lambda^2 \int^{x^+}
 \frac{dx^{+\prime}}{A(x^{+\prime})} }\
e^{-\lambda^2 \int^{x^-} \frac{dx^{-\prime}}{B(x^{-\prime})} } \ ,
\label{33}
\end{equation}
where $u$ is a (negative) constant.
Now inserting (\ref{33}) and (\ref{30}) into (\ref{1cond})
shows that for given
$A(x^+)$ and $B(x^-)$, the conformal factor takes the form of a product
\begin{equation}
e^{2\rho} = u \ \frac{e^{-\lambda^2 \int^{x^+}
 \frac{dx^{+\prime}}{A(x^{+\prime})} }\ }{A(x^+)} \ \frac{
e^{-\lambda^2 \int^{x^-} \frac{dx^{-\prime}}{B(x^{-\prime})} }} { B(x^-)} \ .
\label{60}
\end{equation}
 But within the conformal gauge, a change of coordinates
$\tilde{x}^+(x^+)$ and $\tilde{x}^-(x^-)$ can always be made and the
 conformal factor then undergoes the
transformation $e^{ 2\rho} \rightarrow e^{ 2\rho} \frac{dx^+}{d\tilde{x}^+}
 \frac{dx^-}{d\tilde{x}^-} $.
In view of (\ref{60}), it is clear that the various  $A(x^+)$ and $B(x^-)$
 simply correspond
to different choices of coordinates.  In particular, one can
pick the Kruskal gauge by  arranging for
$\rho$ to vanish. In that case, (\ref{29}) yields
 \begin{equation}
\phi=a_1x^+ x^-+a_2 x^+ + a_3 x^- + a_4 \ ,
\end{equation}
where $a_i$'s are constants. Inserting this into (\ref{1cond}), one finds
$a_1=-\lambda^2$ and $a_2a_3+ \lambda^2a_4= {C}/{4} $. By an appropriate
  translation along
$x^+$ and $x^-$,
  $a_2$ and $a_3$ can always be set to zero. A general solution
is thus
\begin{equation}
\rho=0 \ \ \ \ \ \ \ \ \ \ \phi=\frac{C}{4\lambda^2}-\lambda^2 x^-x^+ \ .
\label{bhs} \end{equation}
When reverting back to the initial variables, this is (as expected) the CGHS
\cite{cghs}
black hole  solution
\begin{equation}
ds^2=-e^{ 2\bar{\rho}}\:dx^+dx^- \label{43}
\end{equation}
with $ \bar{\rho}=\bar{\phi}$ and $e^{-2\bar{\rho}}=\frac{C}{4\lambda^2}-
\lambda^2 x^-x^+ $,
with an ADM
mass $M\equiv{C}/{4\lambda}$. That this quantity is the ADM mass can
be understood from an analysis of the Hamiltonian \cite{kuns3}.
 A special case is the ``linear
dilaton vacuum solution" (LDV) which is obtained for $C=0$.
Clearly, a different choice of coordinates will lead to
$A(x^+)=1$ and $B(x^-)=-1$.  From (\ref{30}) one recognizes that these are the
static solutions which were obtained in Ref. \cite{kuns2}
to prove the validity of Birkhoff's theorem in two-dimensional
dilaton gravity.
Evaluating the observable (\ref{ob}) for the CGHS solution
derived above just yields the constant $C$.

\subsection{$S_0$ and the entropy}

In the following we evaluate Hamilton's principal function (\ref{s0})
on a spacetime that is recovered from the Hamilton-Jacobi equation (\ref{hje}).
This is interesting because one can discuss a relation to the entropy
of the black hole.
 In the conformal gauge we have with (\ref{fd}) and (\ref{rd})
\begin{equation}
S_0=\int dx\: \left [-2\dot{\phi} +  \phi^\prime
\ln\left( \frac{ \phi^\prime+\dot{\phi} }{ \phi^\prime-\dot{\phi} }
 \right)\right] \ ,\label{s0c}
\end{equation}
which only depends on the dilaton field.
 For the CGHS solution (\ref{bhs}), (\ref{s0c}) reads
\begin{equation}
S_0=2\lambda^2 \int dx\: \left [ 2t + x
\ln\left( \frac{ x-t }{x+t }
 \right)\right]   \label{s0cghs}
\end{equation}
Interestingly, the real part of (\ref{s0cghs}) is vanishing
 (contributions from interior
and exterior cancel one another
separately for $x<0$ and $x>0$).
The imaginary part has been interpreted in \cite{kuns3} as being
proportional to the entropy of the black hole. We note from
(\ref{s0c}) that
\begin{equation}
\mbox{Im}S_0= \int_{-\infty}^{\infty}dx\ \phi'
     \mbox{Im}\left(\ln\frac{2\phi'-Q}{2\phi'+Q}\right). \label{bh}
 \end{equation}
Since this yields a non-vanishing value only if
$4\phi'^2-Q^2\leq0$, i.e., inside the horizon, one has
\begin{equation} \mbox{Im}S_0=\pi \int_{-t}^t dx\ \phi' =0.
\end{equation}
How can this result be reconciled with the non-vanishing entropy
of the hole? The contribution to $\mbox{Im}S_0$ from each  horizon is
\begin{equation}
\pi\phi\vert_{horizon}=\frac{C\pi}{4\lambda^2} \ ,\label {ima}
\end{equation}
which is one quarter of the hole's entropy. The result in
\cite{kuns3} will thus be obtained if only the crossing point
of the hypersurface $t=constant>0$ with the upper branch
of the future  horizon is taken into account\footnote{One
of us (J.-G. D.) is grateful to G. Kunstatter for discussions
about this point.}. All this
is independent of the specific hypersurface chosen.
However,  crossing to the inside of
the line of singularity $\phi=0$ should not be done, as it implies an
imaginary value for the physical field $\bar{\phi}$. Thus, one should in
principle keep
$\lambda t<t_c \equiv \sqrt{M/\lambda^3 }$. For $t>t_c$,
integrating for positive $x$ down to the line of singularity $\phi=0$ would
give a single horizon crossing, with
 now $\mbox{Im}S_0= {C\pi}/{4\lambda^2}$, but of
course with a non-vanishing real part.

To get an idea of how general the vanishing of the real part
 $\mbox{Re}S_0$ is,
 we consider
a different model. For four-dimensional spherically symmetric gravity,
where the angular part is frozen, the corresponding two-dimensional dilaton
gravity action is similar to (\ref{sk}), but with a modified
potential $4\lambda^2 \rightarrow 1/\sqrt{2\phi}$. The
associated black hole solution of mass $m$ is \cite{kuns3}
\begin{eqnarray}
\phi&=& \frac{1}{2} r^2 \nonumber \\ \relax
ds^2&=&r[-(1-2m/r)dt^2+(1-2m/r)^{-1})dr^2] \nonumber \\ \relax
&=& -2m e^{-\frac{r}{2m}}d\bar{u}d\bar{v}\ .
\label{ssg}
\end{eqnarray}
In the last equality for the metric, we made use of the usual Kruskal
coordinates,
$\bar{u}=-4me^{-u/{4m}}$ and $\bar{v}=4me^{v/{4m}}$,
where $u=t-r^*$ and $v=t+r^*$
and $r^*=r+2m\ln |r/{2m}-1|$ is the usual tortoise coordinate.
The expression for $S_0$ evaluated on a slice of constant Kruskal time $T$,
defined  from the null coordinates through $\bar{u}\bar{v}=T^2-X^2$, is
(we note that \cite{kuns1} gives
the solution to the Hamilton-Jacobi equation
(\ref{hje}) for {\em arbitrary} potentials)
\begin{equation}
S_0=\frac{1}{2}  \int dX\: e^{-\frac{r}{2m}} \left [ 2T + X
\ln\left( \frac{ X-T }{X+T }
 \right)\right] \ ,  \label{s0ssg}
\end{equation}
in close resemblance with (\ref{s0cghs}). However, the additional
exponential factor
makes $\mbox{Re}S_0$ non-vanishing, as can easily be checked numerically.

\section{Matter evolution} \label{me}
\subsection{Recovery of the Schr\"odinger Equation}
At the next order, $G^0$, we have
\begin{equation}
-\frac{1}{2} \left(\frac{\delta S_0}{\delta \phi}  \frac{\delta S_1}
{\delta \rho} +
\frac{\delta S_0}{\delta \rho}  \frac{\delta S_1}{\delta \phi}\right)-
\frac{1}{2i}\frac{\delta^2
 S_0}{\delta \rho\delta \phi }+\frac{1}{2} \left( e^{-iS_1}
\frac{ -\delta^2 (e^{iS_1})}{\delta  f ^2}
+{f^\prime}^2\right)=0.
\label{G0}
\end{equation}
We now follow the four-dimensional case \cite{kiefer1} and set
 $e^{iS_1}=D^{-1}[\rho,\phi] \chi [\rho,\phi,f]$ so that (\ref{G0}) can
be written as
\begin{eqnarray} & &
\frac{-iD^{-1}}{2} \left(  \frac{\delta S_0}{\delta \phi}  \frac{\delta
D}{\delta \rho} +
\frac{\delta S_0}{\delta \rho}  \frac{\delta D}{\delta \phi}\right)-
\frac{1}{2i}\frac{\delta^2
 S_0}{\delta \rho\delta \phi } \nonumber \\ \relax   & & \;\; +
\chi^{-1}\frac{1}{2}
\left[  i \frac{\delta S_0}{\delta \phi}
\frac{\delta \chi}{\delta \rho} + i
\frac{\delta S_0}{\delta \rho}  \frac{\delta \chi}{\delta \phi}
-   \frac{\delta^2
\chi}{\delta f^2 }+{f^\prime}^2 \chi \right]=0. \label{dc}
\end{eqnarray}
This can be further simplified if $D$ is assumed to obey the equation
\begin{equation}
\frac{\delta^2S_0}{\delta\rho\delta\phi}D-\frac{\delta S_0}{\delta\phi}
\frac{\delta D}{\delta\rho}
 -\frac{\delta S_0}{\delta\rho}\frac{\delta D}{\delta\phi}=0. \label{d}
\end{equation}
 The first term in (\ref{d}) is formally infinite since it involves
functional derivatives at the same point. Some authors have argued that,
 after suitable regularization,
 it can be ignored \cite{hori,dw} and that one may solve by $D=1$ as if
there were no matter. However, as already mentioned above,
this can at best be considered ad hoc, but
for the purpose of the present paper it is not necessary to
resolve this issue.

 We are thus left with
\begin{equation}
-\frac{i}{2} \left( \frac{\delta S_0}{\delta \phi} \frac{\delta \chi}{\delta
\rho}+
 \frac{\delta S_0}{\delta \rho} \frac{\delta \chi}{\delta \phi}
\right)={\cal{H}}_m \chi   \ ,
\label{tse}
\end{equation}
where ${\cal{H}}_m$ is the matter Hamiltonian density,
${\cal H}_m \equiv\frac{1}{2} (- \frac{\delta^2  }{\delta f^2 }+{f^\prime}^2)$.
Use of (\ref{tse}) may appeal to either the first or second equalities in
(\ref{fd}) and (\ref{rd}).
In the next section, we make use of the second equalities  to
 discuss a difficulty in identifying the LHS of (\ref{tse}) as a
Tomonaga-Schwinger type time.
Here, we derive results for a classical background, making
use of the first equalities.
In the conformal gauge ($\sigma =1$ and $\xi=0$),
(\ref{tse}) takes the simple form
\begin{equation}
i\left( \dot{\rho}\frac{\delta \chi}{\delta \rho}+
  \dot{\phi}\frac{\delta \chi}{\delta \phi}\right)={\cal{H}}_m \chi
\ ,
\label{tse2}
\end{equation}
which can be immediately integrated to yield
 \begin{equation}
i\frac{\partial\chi}{\partial t}=H_m \chi
\label{tse3}
\end{equation}
with $H_m\equiv\int dx{\cal H}_m$.
This is the functional Schr\"odinger equation for a free scalar field
on a flat background. The only information about the gravitational
fields occurs in the definition of time through $\rho$ and $\phi$.
Note that the use of the original, physical, field variables
yields an equation whose form is equivalent to that of (\ref{tse2}):
\begin{equation}
i \left(\dot{\bar{\rho}}\frac{\delta \chi}{\delta \bar{\rho}}+
 \dot{\bar{\phi}}\frac{\delta \chi}{\delta \bar{\phi}}\right)=
{\cal{H}}_m \chi \ .
\label{tse4}
\end{equation}
The reason for this formal similarity is of course the conformal
coupling of the matter field in two dimensions.
Integrating (\ref{tse4}) gives again the  free evolution (\ref{tse3}),
but this time on the black hole background (\ref{43}).

\subsection{Consistency}

As it has been shown recently \cite{kiefer2}, replacing the
left-hand side of (\ref{tse}) with
a functional derivative $d/{\delta \tau(x)}$ is not consistent, since
the object on the left-hand side does not commute
 for two different space points.
 In fact,
it was found that this commutator on both sides of (\ref{tse}) amounts to the
invariance of $\chi$
under spatial diffeomorphisms. In our simple two-dimensional model,
this can easily be shown explicitly. Denote the vector fields
-- one vector field at each space point $x$ --
acting on the left-hand side of (\ref{tse}) by
\begin{equation} v(x)\equiv-\frac{1}{2}\left(\frac{V_G}{Q}
\frac{\delta}{\delta\rho}
 +Q\frac{\delta}{\delta\phi}\right), \end{equation}
and consider their smeared-out version \cite{kiefer2}
\[ v^N\equiv \int dx N(x)v(x). \]
Consistency of (\ref{tse}) then demands that
\begin{equation} [v^N,v^M]\chi=[H_m^M,H_m^N]\chi, \end{equation}
where, of course, $H_m^N\equiv \int dxN(x){\cal H}_m(x)$.
Explicit calculation yields for the left-hand side
\[ [v^N,v^M]\chi =\int dx(NM'-MN')\left(\phi'\frac{\delta}{\delta\phi}
 +\rho'\frac{\delta}{\delta\rho}-\frac{d}{dx}\frac{\delta}
{\delta\rho}\right)\chi, \]
while for the right-hand side one finds
\[ [H_m^M,H_m^N]\chi =\int dx(MN'-NM')f'\frac{\delta\chi}{\delta f}.\]
This, however, is nothing but the momentum constraint (\ref{mom})
in this order of approximation. As has been emphasized in
\cite{kiefer2}, equations like (\ref{tse}) have to be properly
interpreted in their integrated form, as a functional
Schr\"odinger equation, after a specific choice for lapse and shift
has been made, so as to recover quantum field theory on a
specific family of spacetimes. This does not, however, mean that
the local form of the Schr\"odinger equation (the Tomonaga-Schwinger
equations) is useless, since it may be used for formal
considerations, for example the discussion of the correction terms
in Section~4.5.

\subsection{Hawking radiation in a collapsing geometry}
\label{collapse}
We now show how to recover Hawking radiation from solutions
to (\ref{tse3}). This is complementary to the standard analysis
employing the calculation of Bogolubov coefficients
\cite{GN} in that the use of the Schr\"odinger picture brings out
interesting new aspects.
Rather than an eternal black hole (\ref{43}), we first consider  the
`collapsing' background of \cite{cghs}.
The formally analogous case of the eternal black hole (\ref{43})
 will be treated
in the next subsection.
Although (\ref{tse3}) was obtained in the context of  the
eternal black hole,
it is clear from the point of view of quantum field theory
 on curved geometries and
 from  conformal invariance that  it also governs the evolution on an
arbitrary two-dimensional background, where the variable $t$ stands for the
 time variable of the conformally flat metric.
The collapsing spacetime we are interested in  can be obtained from
the action (\ref{cghs}) by assuming that
 a left moving shock wave of
classical $f$ - matter is imparted (for example)
 at $\lambda x^+=1$, producing a stress tensor
$\frac{1}{2}\partial_+f \partial_+f=\lambda M \delta (\lambda x^+-1)$,
 so as to form the black hole. In the Kruskal gauge,
where $\bar{\rho}=\bar{\phi}$,
one then has the background
$ds^2=-e^{2\bar{\rho}}dx^+dx^-$ with
\begin{equation}
e^{-2\bar{\rho}}=e^{-2\bar{\phi}}=\frac{M}{\lambda}(1-\lambda x^+)
\:\Theta ( \lambda x^+ -1)-\lambda^2x^+x^-
\label{cghsbg}
\end{equation}
where $\Theta(x)$ is the usual step function.
The
corresponding Penrose diagram is presented in Figure~1.
\begin{figure}
\vspace{-.55truein}
\hbox {\hspace{2.65cm} \resizebox{9cm}{!}{
\includegraphics*[40mm,60mm][180mm,186mm]{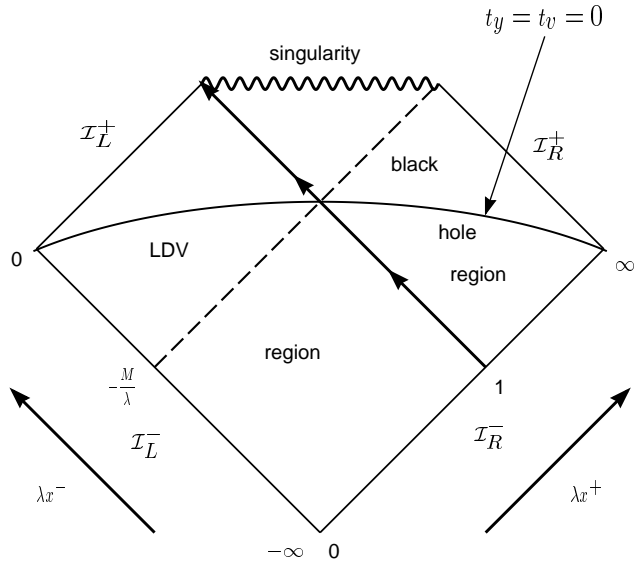}} }
\hbox{\hspace{2.0cm} \parbox{10cm}{\caption{{\small
Penrose diagram for the CGHS collapsing black hole. Shock vave (horizon) are
indicated by the arrowed (dashed) line.
The overlapping slices $t_y=t_v=0$, where the matter state  evolving from the
past is compared with the one escaping to $I_R^+$, are also indicated.
} }}}
\vspace{-.15truein}
\label{fig1}
\end{figure}
One might wonder how a nontrivial effect such as Hawking radiation
can emerge from (\ref{tse3}), since this equation has the form of
a free Schr\"odinger equation. The reason lies in the proper
formulation of {\em boundary conditions}. The idea is to start
from the vacuum state for the scalar field in the absence of a
black hole at early times (the ``linear dilaton vacuum" region,
LDV), let it then evolve
according to (\ref{tse3}) and compare it with the vacuum solution
in the presence of a black hole at late times.
The notion ``vacuum state" is here defined with respect to
``inertial" coordinates, i.e., coordinates which exhibit
explicitly the asymptotic flatness of the metric.
For (\ref{cghsbg}), such coordinates are respectively
\begin{equation}
\begin{array}{lll}
\lambda x^+=e^{\lambda y^+}& \lambda x^-=-\frac{M}{\lambda}
e^{-\lambda y^-}& LDV\  region \nonumber \\ \relax
 & & \nonumber \\ \relax
\lambda x^+=1+\frac{\lambda}{M}e^{\lambda v^+}& \lambda
x^-=-\frac{M}{\lambda}-
e^{-\lambda v^-}& black\  hole\  region\ .
\end{array} \label{coor}
\end{equation}
{}From these lightcone coordinates,
 timelike and spacelike directions are defined as
usual via $y^\pm=t_y\pm y$ and $v^\pm=t_v\pm v$.
Note that $v^\pm$ only covers the region
 above the shock wave, $\lambda x^+\geq 1$, and outside the horizon,
$\lambda x^-\lambda \leq -{M}/{\lambda}$.
The comparison of the vacuum states will be made on the
(overlapping) slices $t_y=t_v=0$. In order to obtain such overlapping
slices, the coordinate transformation chosen in (\ref{coor}) is
slightly different from a similar one used in the literature,
see \cite{stro}. As
 $v\in(-\infty,\infty)$ on the
half-line $y \geq 0 $, we must demand the boundary condition
$f(v)\to 0$ as $v\to\pm\infty$ for the field $f$ because otherwise
one would not expect the Gaussian functionals which play the
role of vacuum states (see below) to converge. In the
coordinate $y$ this condition means that $f$ has to {\em vanish}
at the origin $y=0$. This turns out to be crucial for the
following discussion.

It is convenient to consider the Fourier transform of the fields.
Because of the boundary condition we have
\begin{equation} f(y)=\sqrt{\frac{2}{\pi}} \int_0^{\infty} dk\ \tilde{f}(k)
    \sin ky, \label{fx} \end{equation}
where $\tilde{f}(k)=\tilde{f}^*(k)$ (reality of $f(y)$). In the following
we shall omit the tilde for simplicity and denote the Fourier
transform by $f(k)$. It follows from (\ref{fx})  that
\begin{equation} \frac{\delta}{\delta f(y)}= \sqrt{\frac{2}{\pi}}
    \int_0^{\infty}dk\ \sin ky\frac{\delta}{\delta f(k)},\hspace{1.5cm}
     \frac{\delta f(k')}{\delta f(k)}=\delta(k-k'). \end{equation}
The functional Schr\"odinger equation (\ref{tse3})  then reads
\begin{equation} \frac{1}{2}\int_0^{\infty}dk\ \left(-
\frac{\delta^2}{\delta f^2(k)}
    +k^2f^2(k)\right)\chi =i  \frac{\partial\chi}{\partial t_y }\ .
\end{equation}
We shall choose the ground state solution of this equation.
In the present case this can be represented by a Gaussian functional
(see for example \cite{Ja}), and we have
\begin{equation} \chi_0[f,t_y) =N\exp\left(-\frac{1}{2}\int_0^{\infty}
    dk\ kf^2(k) \ -i  E_0 t_y \right). \label{gr} \end{equation}
The ground state energy $E_0$ is of course divergent and must
be regularized. We shall choose the ground state solution
(\ref{gr}) as the {\em initial condition} for the Schr\"odinger equation
in the `future' variables $(t_v,v)$. This equation reads
\begin{equation} \frac{1}{2}\int_{-\infty}^{\infty}dv\
    \left(-\frac{\delta^2}{\delta f^2(v)}+ \left[\frac{\partial f}{\partial
    v}\right]^2\right)\chi_b= i  \frac{\partial\chi_b}{\partial t_v},
    \label{til} \end{equation}
where the subscript ``$b$" refers to the ``black hole region".
 The important difference
to above is that $f(v)$ has no restriction at $v=0$.
Therefore the Fourier transform is simply given by
\begin{equation} f(v)= \int_{-\infty}^{\infty}\frac{dk}{\sqrt{2\pi}}\
    g(k)e^{ikv}, \label{gk} \end{equation}
which has to be contrasted with (\ref{fx}). In that case,
\begin{equation} \frac{\delta}{\delta f(v)}= \int_{-\infty}^{\infty}
    \frac{dk}{\sqrt{2\pi}}\ e^{-ikv}\frac{\delta}{\delta g(k)},\hspace{1.5cm}
    \frac{\delta g(k')}{\delta g(k)}=\delta(k-k'). \end{equation}
Thus, Eq. (\ref{til}) becomes
\begin{equation} \frac{1}{2}\int_{-\infty}^{\infty}dk\ \left(-\frac{\delta^2}
    {\delta g(k)\delta g^*(k)} +k^2\vert g(k)\vert^2\right)\chi_b
   =i\frac{\partial}{\partial t_v}\chi_b. \label{schg} \end{equation}
Note that $g(k)$ is {\em complex}, whereas $f(k)$ is real.
At $t_y=t_v=0$, the space slices overlap and the states may be compared.
The ground state solution to (\ref{schg}) at $t_v=0$ reads
\begin{equation} \chi_{b,0}= N\exp\left(-\int_{-\infty}^{\infty}
    dk\ \vert k\vert\ \vert g(k)\vert^2\right). \label{vac} \end{equation}
This is of course different from the previous ground state solution
(\ref{gr}). To use (\ref{schg}) we have to rewrite the ground state
(\ref{gr}) in terms of the field $g(k)$. We thus
write a Bogolubov relation
\begin{equation} f(k)= \int_{-\infty}^{\infty}dl\ \alpha(k,l)g(l)
\hspace{1cm} (k>0)
\label{bogor} \end{equation}
and therefore (using (\ref{fx}) and (\ref{gk}))
\[ \int_0^{\infty}dy\ f(y)\sin ky= \frac{1}{2}\int_{-\infty}
   ^{\infty}dl\ \alpha(k,l)\int_{-\infty}^{\infty}dv\ f(v)e^{-ilv}. \]
One can see by explicit calculation that this can be
satisfied by the ansatz
\begin{equation} \alpha(k,l) =\frac{1}{\pi}\int_0^{\infty} dy\ \sin ky\:
    e^{ilv(y)}, \end{equation}
with $v(y)$ being the coordinate transformation (\ref{coor})
at $t_y=t_v=0$.
This leads to
\begin{eqnarray} \alpha(k,l) &=& \frac{1}{\pi}\int_0^{\infty}
     dy\ \sin ky \: e^{\frac{il}{\lambda} \ln\frac{M}{\lambda}
(e^{\lambda y }-1)} \nonumber\\
        &\approx & \frac{1}{\pi}\left (\frac{M}{\lambda}\right )^
{\frac{il}{\lambda} }
\int_0^{\infty}  dy\ \sin ky\:(\lambda y)^{\frac{il}{\lambda}}
\nonumber \\ \relax
     &\approx & \frac{1}{\pi\lambda}
\left (\frac{M}{\lambda}\right )^{\frac{il}{\lambda} }
\Gamma\left(1+\frac{il}{\lambda}\right)
     \cosh\frac{\pi l}{2\lambda}\left\vert\frac{k}{\lambda}\right\vert
     ^{-1-\frac{il}{\lambda}}\ .\label{alpha}
\end{eqnarray}
In the second step of (\ref{alpha}), the standard approximation
$(e^{\lambda y}-1) \approx \lambda y$ was made, picking up the
dominant contribution in
the neighborhood of the horizon \cite{GN,Haw}, and
 3.763.1 of \cite{GR} was used in the last step.
Taking the state (\ref{gr}) at $t_y=0$ and expressing it with respect
to the field $g(k)$, one finds
\begin{eqnarray} \chi_0 &=& N\exp\left(-\frac{1}{2}\int_0^{\infty}
      dk\ k\int_{-\infty}^{\infty}dp\: dp'\ \alpha(k,p)\alpha(k,p')
     g(p)g(p')\right)\nonumber\\
  &=& N\exp\left(-\int_{-\infty}^{\infty}dp\ p\coth
     \frac{\pi p}{2\lambda}\vert g(p)\vert^2\right)\ ,
 \label{coth} \end{eqnarray}
which is independent of the black hole mass $M$.
Because of the analogy to the Rindler case, namely the ``loss" of the
information on half the space slice behind the horizon,
this is in accordance
with a similar result found in \cite{FHM} in the context of
accelerated observers and the Unruh radiation.

To solve (\ref{schg}) with the initial condition (\ref{coth}),
it is again appropriate to make a Gaussian ansatz:
\begin{equation} \chi_b= N(t_v)\exp\left(-\int_{-\infty}^{\infty}
    dp\ p\: \Omega(p,t_v)\vert g(p)\vert^2\right). \label{chi1} \end{equation}
Inserting this into (\ref{schg}) leads to
\begin{equation} -\frac{i}{p}\frac{\partial\Omega}{\partial t_v}
   =1-\Omega^2. \label{dOm} \end{equation}
The solution of this equation with the initial condition
$\Omega(p,0)=\coth(\pi p/2\lambda)$ is simply given by
\begin{equation} \Omega(p,t_v)= \coth\left(\frac{\pi p}{2\lambda}
+ipt_v\right).
    \label{Om} \end{equation}
We now consider the number operator of the mode with wave number $k$, which
is associated with the vacuum state (\ref{vac}) in the presence of the
hole. Calculating its expectation value
with respect to the state (\ref{chi1}) -- the time-developed state of
the ``free" vacuum -- one finds
\begin{equation} \langle n(k)\rangle = \frac{(\Omega_R-1)^2+\Omega_I^2}
     {4\Omega_R}, \end{equation}
where $\Omega_R$ ($\Omega_I$) denotes the real (imaginary) part
of $\Omega$. At $t_v=0$ this is given by
\begin{equation} \langle n(k)\rangle =\frac{1}{e^{
\frac{2\pi\vert k\vert}{\lambda}}
   -1}, \label{num} \end{equation}
i.e. a Planck distribution with the temperature $\lambda/2\pi$,
as expected. Employing the differential equation (\ref{dOm})
one recognizes that $d\langle n\rangle/dt_v=0$, i.e. the Planck spectrum
is conserved in time.

It is important to emphasize that the Planck spectrum has been
recovered from a {\em pure} quantum state. This means that there
exist other operators than the number operator, from which it is
possible to recognize explicitly the difference to a thermal state.
Note also that the density matrix of a canonical ensemble of oscillators
contains some extra-term besides the $coth$-term showing up in
(\ref{coth}).

\subsection{Eternal black hole}

Let us now consider how Hawking radiation may arise in the eternal black hole
background (\ref{43}) (which due to its time symmetry would be more
properly called a ``black-and-white hole").
The derivation in fact closely parallels the one of the collapsing hole, so
a brief presentation will suffice.
To describe the radiation escaping to the RHS of the black hole, we introduce
the coordinates
\begin{eqnarray}
\lambda t &=& e^{\lambda \tilde{x} } \sinh \lambda \tilde{t}
 \nonumber \\ \relax
\lambda x &=& e^{\lambda \tilde{x} } \cosh \lambda \tilde{t} \ ,
\label{ttxt}
\end{eqnarray}
which cover the wedge $|x|>t$ of the manifold.
The metric (\ref{43}) then takes the form
$ds^2=(1+M/\lambda e^{-2\lambda \tilde{x}})^
{-1}(-d\tilde{t}^2+d\tilde{x}^2)$
 which is thus asymptotically flat far from the black hole.
It is clear that the slices $\tilde{t}=0$ and $t=0$ coincide,
so that states evolving in $(t,x)$
and $(\tilde{t},\tilde{x})$ may be compared there.
As before, we then have the feature that  the range $\tilde{x}
\in (-\infty, \infty)$
only covers the half-line $x\geq 0$. As a result, requiring
the matter field configuration $f(\tilde{x})$ to vanish at
spatial infinity $\tilde{x}
\rightarrow \pm \infty$ will translate into $f(x=0)=0$.
Note the analogy of (\ref{ttxt}) to the transformation between
Minkowski coordinates and Rindler coordinates, from which it is
immediately clear that the Hawking temperature here is $T_{BH}
=\lambda/2\pi$, independent of the mass. This has recently also been
emphasized in \cite{CM}.

We first determine the vacuum state $\bar{\chi}_0[f(k),t]$ which is a
 solution of the
Schr\"o\-dinger equation (\ref{tse3}) in $(t,x)$ coordinates
 ($f(k)$ being the Fourier tranform of $f(x) $ as in (\ref{fx}).
Although the metric is not
asymptotically flat in those coordinates, a vacuum state may still be defined.
As we do not wish to involve arbitrarily short distance physics, we choose
for initial Cauchy surface a slice $t<0$, but $t>-\sqrt{M/\lambda^3}$.
Following
essentially the same steps as before,  $\bar{\chi}_0[f(k),t]$ is just the
 state (\ref{gr}) (with the change $t_y\rightarrow t$).
The state will then evolve up to $t>0$.
For the evolution in  terms of the  $(\tilde{t},\tilde{x})$ coordinates,
the transformation (\ref{ttxt}) is conformal, so (\ref{tse3})
still applies and the
corresponding vacuum state at $\tilde{t}=0$,
  $\bar{\chi}_{R,0}[g(k),\tilde{t}=0]$, is thus the same as (\ref{vac}).
To determine how $\bar{\chi}_0[f(k),t=0]$
reads in terms of $g(k)$, we write a Bogolubov relation as in (\ref{bogor}),
and find the exact result
\begin{equation}
\alpha(k,l) =\frac{1}{\pi\lambda} \Gamma\left(1+\frac{il}{\lambda}\right)
     \cosh\frac{\pi l}{2\lambda}\left\vert\frac{k}{\lambda}\right\vert
     ^{-1-\frac{il}{\lambda}} \ .
\end{equation}
Then, following steps analogous to those leading to (\ref{num})
from (\ref{coth}),
we obtain
that the state thus evolved according to time $\tilde{t}$,
$\bar{\chi}_R[g(k),\tilde{t}]$,
has indeed a thermal content in the sense of (\ref{num}).

We also note that the non-vanishing of $\mbox{Im}S_0$ (compare (\ref{bh}))
is easily understood for the eternal hole. The integral in
(\ref{bh}) has to be performed with respect to $\tilde{x}$ which ranges
from $-\infty$ to $\infty$ along the half-line originating from the
origin in the $(t,x)$-diagram. Since it has only one point in common
with the horizon, one directly finds the desired result (\ref{ima}).

\subsection{Corrections to the Schr\"odinger Equation}

We now proceed to the next order of the semiclassical
approximation, ${\cal O}(G)$, to show how the functional
Schr\"odinger equation is modified by quantum gravitational
corrections \cite{kiefer1,KS}. One first obtains an equation involving
$S_2$,
\begin{eqnarray} & & -\frac{1}{2}\left(\frac{\delta S_0}{\delta\phi}
 \frac{\delta S_2}{\delta\rho}+ \rho\leftrightarrow\phi\right)
 -\frac{1}{2}\frac{\delta S_1}{\delta\phi}\frac{\delta S_1}{\delta\rho}
 +\frac{i}{2}\frac{\delta^2S_1}{\delta\phi\delta\rho} \nonumber\\
 & & \; +\frac{\delta S_1}{\delta f}\frac{\delta S_2}{\delta f}
 -\frac{i}{2}\frac{\delta^2S_2}{\delta f^2}=0. \end{eqnarray}
As in the general, four-dimensional, case, this can be greatly
simplified by rewriting $S_2=\sigma _2[\rho,\phi]+ \eta[\rho,\phi,f]$ and
demanding, in analogy to (\ref{d}), an equation for $\sigma _2$ such that the
equations simplify. Since these steps are in full analogy to the
general case, we shall be very brief here. Introducing a
functional
\begin{equation} \psi=\chi e^{i\eta G}, \end{equation}
one finds for $\psi$ the ``corrected Schr\"odinger equation"
\begin{eqnarray} & & -\frac{i}{2}\left(\frac{V_G}{Q}
\frac{\delta\psi}{\delta\rho}
 +Q\frac{\delta\psi}{\delta\phi}\right) ={\cal H}_m\psi +
 \nonumber\\
& & \; \frac{G}{\chi}\left(-\frac{1}{2D}\frac{\delta D}{\delta\rho}
   \frac{\delta\chi}{\delta\phi} +\phi\leftrightarrow\rho
  +\frac{1}{2}\frac{\delta^2\chi}{\delta\phi\delta\rho}\right)\psi.
 \end{eqnarray}
If $D=1$, the only correction term reads
\begin{equation} \Delta{\cal H}_m\psi\equiv \frac{G}{2\chi}
\frac{\delta^2\chi}
 {\delta\phi\delta\rho}\psi, \end{equation}
which, of course, involves second order derivatives with respect to
the gravitational degrees of freedom. If $\chi$ is a solution to
 (\ref{tse3}),
one would expect that it separates in $f$ and the gravitational
variables. In this case the correction term
would only yield a contribution
to the phase, which should not be important.
However, as the analysis in \cite{KMS} suggests, this term should
become relevant if the mass of the hole approaches the Planck mass.
To evaluate the correction terms properly, one has first to employ
a careful regularization, since second functional derivatives
are involved.
Since this is beyond the scope of this paper, it will be
relegated to a future publication.

\section{Decoherence} \label{deco}

The derivation of the functional Schr\"odinger equations (\ref{tse3})
 was achieved through the use
of a particular WKB state for the gravitational sector.
Namely, the choice $\Psi\approx e^{iS_0}\chi$ and $\sigma >0$
was made (although
the same equation can equally be obtained  with
$\Psi\approx e^{-iS_0}\chi$ and $\sigma <0$).
But since the fundamental equations
which are assumed here, (\ref{mc}) and (\ref{wdw}), are {\em linear},
one would,
however, expect that arbitrary superpositions of WKB-type
states occur. This is suggested also by the real nature of the
Wheeler-DeWitt equation, from which it would seem artificial to
choose a special complex solution. A more natural state
would thus be
\begin{equation} \Psi\approx e^{iS_0}\bar{\chi}_R +
e^{-iS_0}\bar{\chi}^*_R\ , \label{bwh} \end{equation}
where $\bar{\chi}_R$ is the state which evolves out of the
eternal black hole
geometry and reads
\begin{equation} \bar{\chi}_R=N(\tilde{t})
\exp\left(-\int_{-\infty}^{\infty}
    dk\ k \: \coth \left(\frac{\pi k}{2\lambda}+
ik\tilde{t} \right)\vert g(k)\vert^2\right) \label{chit} \end{equation}
The state (\ref{bwh})
 could naively be called a ``superposition of a black hole
with a white hole". Such states arise, for example, in the
analysis of the gravitational collapse of a dust shell with
a sensible boundary condition for the wave function \cite{Haj}.
Another possibility would be
\begin{equation} \Psi\approx e^{iS_0}\bar{\chi}_R +
e^{-iS_0^{(0)}}\bar{\chi}_0, \label{bnh} \end{equation}
where the second component refers to the linear dilaton
vacuum (thus $C=0$), i.e.
\begin{equation} \bar{\chi}_0 =N_0(\tilde{t})\exp\left(-\int_{-\infty}
     ^{\infty} dk\ \vert k\vert \ \vert g(k)\vert^2\right). \end{equation}
 One can interpret the state (\ref{bnh}) as a
superposition of a black hole with no hole.
States of this kind
have in fact been discussed extensively in QED and quantum
cosmology \cite{Ki2}.
It has been demonstrated there that the presence of a huge number
of ``irrelevant" degrees of freedom may cause decoherence of such states.
In the case of QED these may be charged matter states, while in quantum
cosmology these may be general matter states or states describing
gravitational waves. Because states such as (\ref{bwh}) are found
in a Born-Oppenheimer approximation, the decoherence of the
various components is nothing but an expression of spontaneous
symmetry breaking as it happens, for example, in the case
of chiral molecules \cite{Ze}.
What could possibly play the role of the decohering agent
in the present case? From the discussion in the preceding section
one would expect that Hawking radiation may be able to suppress
interferences between the separate WKB-states. This is in particular
suggested by its irreversible nature and the fact that it provides
a huge entropy capacity.

Writing $\Psi^*\Psi$ as a sum of four terms ($\Psi$ being the
superposition (\ref{bwh})), the degree of decoherence between
the two semiclassical components can be studied from the
following component of the reduced density matrix for the
gravitational sector
\begin{equation} \rho_{\pm} [\rho,\phi] =e^{2iS_0}
\int {\cal D } g \: {\cal D } g^*
  \bar{\chi}_R^2[g,g^{*},\tilde{t})\equiv e^{2iS_0}D_{\pm},
\label{rho} \end{equation}
and its conjugate
 $\rho_\mp=\rho_\pm^*$.
 Because of the quadratic
dependence on the field $g,g^{\dagger}$, the trace in (\ref{rho})
can immediately be evaluated, in full analogy to the QED-case
discussed in \cite{Ki2}. One has, with
$\Omega(k,\tilde{t})$ given in (\ref{Om}),
\begin{eqnarray} D_{\pm}&=& \mbox{det}\frac{\Omega_R}{\Omega}
     =\exp\left(  -\mbox{Tr}\: \ln (1+i\left
 (\frac{\Omega_R}{\Omega_I}\right)\right) \nonumber \\ \relax
   & =&\exp\left(-i\ \mbox{Tr}\frac{\Omega_I}{\Omega_R}
     -\frac{1}{2}\mbox{Tr}\left(\frac{\Omega_I}{\Omega_R}
     \right)^2-\ldots\right). \label{dpm} \end{eqnarray}
The decoherence factor can be be evaluated by noting that the
real and imaginary parts of $\Omega$ are given, respectively,
by the expressions
\begin{eqnarray}
\Omega_R &=& -\frac{\sinh \frac{\pi k}{\lambda}}
{\cos 2k\tilde{t}-\cosh \frac{\pi k}{\lambda} }
\nonumber \\ \relax
\Omega_I &=& \frac{\sin 2k \tilde{t}}{\cos 2k\tilde{t}-\cosh
 \frac{\pi k}{\lambda} }
\label{Ome} \end{eqnarray}
which yields
\begin{equation} \Omega_I =-\frac{\sin 2k\tilde{t}}{\sinh
\frac{\pi k}{\lambda}}
   \Omega_R. \label{ratio}  \end{equation}
One thus has for the exact decoherence factor the expression
  \begin{equation} D_{\pm} =\exp\left(-L\int_0^{\infty}\frac{dk}{2\pi}
   \ln\left[1-i\frac{\sin 2k\tilde{t}}{\sinh\frac{\pi k}
{\lambda}}\right]\right),
  \label{ex} \end{equation}
where $\mbox{Tr}\to L\int dk/2\pi$ was used, and $L$ is a
regularization length. The important part for the amount of decoherence
is its absolute value,
 \begin{equation} \vert D_{\pm}\vert =\exp\left(-L\int_0^{\infty}
\frac{dk}{4\pi}
   \ln\left[1+\frac{\sin^2 2k\tilde{t}}{\sinh^2\frac{\pi k}
{\lambda}}\right]\right).
  \label{ab} \end{equation}
Since this cannot be exactly evaluated, the expansion of the exponent
made in (\ref{dpm}) appears appropriate. Considering the first
real term in the exponent, one finds
\begin{eqnarray} D_{\pm}^{(1)} &\equiv & \exp\left(-\frac{1}{2}
    \mbox{Tr}\left(\frac{\Omega_I}{\Omega_R}\right)^2\right)
 =\exp\left(-\frac{L}{2\pi}\int_0^{\infty}dk\
   \frac{\sin^2 2k\tilde{t}}{\sinh^2\frac{\pi k}{\lambda}}\right)
\nonumber \\ \relax
  &=&\exp\left(-\frac{L\lambda}{4\pi^2}\left[2\lambda \tilde{t}
\coth 2\lambda \tilde{t}
  -1\right]\right), \label{dec} \end{eqnarray}
where 3.986.4 in \cite{GR}
was used in the last step. A finite value for $L$ is obtained,
if the black hole states are put into a box of finite length,
as is frequently done in gedankenexperiments.
In the limit $\tilde{t}\to0$ this goes
to one as it must, since there we matched our state to the one
for the dilaton vacuum which has $\Omega_I=0$. In the limit
$\tilde{t}\to\infty$ this would yield the simple expression
 $\exp(-L\lambda^2 \tilde{t} /2\pi^2)$. One must, however, note that
higher order terms in the expansion (\ref{dpm}) become important
in this limit and one has to resort to the exact expression
(\ref{ab}). Differentiating (\ref{ab}) with respect to $\tilde{t}$
yields an expression which vanishes (due to Riemann's lemma)
in the limit of large $\tilde{t}$. The decoherence factor (\ref{ab})
itself thus approaches a constant. It is convenient to write
\begin{equation} \ln\vert D_{\pm}\vert =-\frac{L\lambda}{4\pi^2}F(y),
 \label{F} \end{equation}
where $y\equiv 2\lambda\tilde{t}/\pi$. Simple numerical analysis
yields a limiting value for $F$ of about $1.65$ which is rapidly
obtained for $y$ bigger than about $10$, as depicted  in Figure~2.
Decoherence thus becomes effective for $L\lambda>1$. This can be physically
interpreted as follows. The dominating mode for the Hawking temperature
$T=\lambda/2\pi$ has a wavelength of the order $\lambda^{-1}$.
If the black hole
is put into a box with radius $L$, interferences are important if
$L$ is of the same order than this wavelength, but they become
suppressed if the box is larger.
\begin{figure}
\hbox {\hspace{2.75cm} \resizebox{9cm}{!}{
\includegraphics*[32mm,110mm][165mm,180mm]{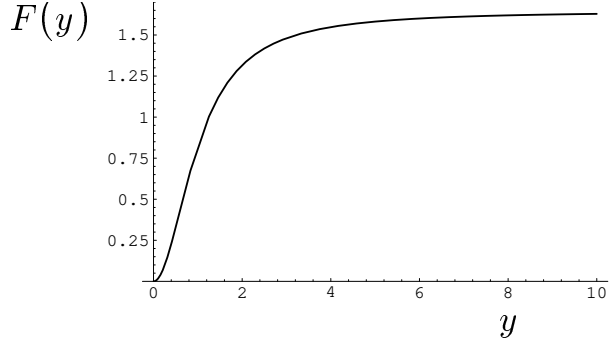}} }
\hbox{\hspace{2.0cm} \parbox{10cm}{\caption{{\small
Plot of $F(y)$, where $y\equiv 2 \l \tilde{t}/\pi$. After a transitory period
of order $\l^{-1}$, the decoherence in Eq. (\ref{F}) reaches a plateau in time
$\tilde{t}$.
} }}}
\label{fig2}
\end{figure}
One can now calculate the analogous expression for the
superposition (\ref{bnh}) of a black hole with no hole.
Instead of (\ref{rho}) one now has to calculate
\begin{equation} \tilde{\rho}_{\pm} [\rho,\phi] =e^{iS_0^{(1)}-iS_0^{(0)}}
\int {\cal D } g \: {\cal D } g^*
  \bar{\chi}_R\bar{\chi}_0^*\equiv \ e^{iS_0^{(1)}
   -iS_0^{(0)}}\ \tilde{ D}_{\pm}, \label{rho1} \end{equation}
Inserting the Gaussian functionals $\bar{\chi}_R$ and $\bar{\chi}_0$
into this
expressions, one
 recognizes that now $\tilde{D}_{\pm}=\mbox{det}[\tilde{\Omega}_R
/(\tilde{\Omega}_R+i\tilde{\Omega}_I)]$, where $\tilde{\Omega}_R=
\Omega_R+1$ and $\tilde{\Omega}_I=\Omega_I$ with $\Omega_R$ and
$\Omega_I$ given by the expressions (\ref{Ome}). In this example
one then finds
\[ \frac{\tilde{\Omega}_I}{\tilde{\Omega}_R}
    =-\frac{\sin 2k\tilde{t}}{e^{\frac{\pi k}{\lambda}}-
\cos 2k\tilde{t}}. \]
This then leads instead of (\ref{F}) to an expression
for the decoherence factor of the form
\[ \ln\vert\tilde{D}_{\pm}\vert
   =-\frac{L\lambda}{4\pi^2}\tilde{F}(y), \]
where
\[ \tilde{F}(y)= \int_0^{\infty}dx\ln\left(1+\frac{\sin^2xy}
   {(e^x-\cos xy)^2}\right) < F(y). \]
One has $\tilde{F}<F$ for all $y$ because
only one component in (\ref{bnh}) carries
Hawking radiation. Again, $\tilde{F}(y)$ approaches a constant
for large $y$,
but -- in contrast to $F$ -- it is not monotonic.
A numerical evaluation yields, for example, the value
$\tilde{F}(15)\approx 0.413$
(compared to $F(15)\approx 1.637$).

Since the state $\bar{\chi}_R$ does not depend on the  black hole mass,
there is of course no decoherence for a superposition corresponding
to different masses, in contrast to the  case of
accelerating detectors\cite{jgd}. This is, however, a peculiarity of the
two-dimensional model. In four dimensions, the Hawking temperature
is inversely proportional to the mass, so one would expect
decoherence for not-to-large masses. Heuristically, one can
take this into account by replacing $\lambda$ in the above expressions
for the states by $(4GM)^{-1}$.
One would thus expect the decoherence factor in four spacetime
dimensions to be of the limiting form for large times
\[ D_{\pm}\approx \exp\left(-\ constant\ \times\left(\frac{L}
  {4GM}\right)^3\right). \]
This expression depends now explicitly on the mass of the hole,
and one recognizes that decoherence is efficient for small masses
where the Hawking radiation is large.
Again, decoherence for a black hole in a box of dimensions $L^3$
would be efficient for the realistic case of the length $L$ being
much bigger than the Schwarzschild radius.

\section{Conclusions}
\label{con}

The central issue in our paper is the investigation of the
semiclassical approximation in the context of two-dimensional
dilaton gravity. We have demonstrated how Hawking radiation
can be  properly understood in the Schr\"odinger representation and
how the validity of the semiclassical approximation can be investigated.
We have shown, in particular, how the correlation between the
Hawking radiation and the black hole can decohere the latter --
the various semiclassical components become dynamically
independent.

We want to conclude in mentioning some of the interesting open problems
which are topics for future research. Hawking
radiation was obtained in our framework from a {\em pure} quantum state
outside the horizon. While the expectation value of the number operator
for the field modes exhibits a perfect Planckian spectrum, there
are of course higher order operators which distinguish this state
from a thermal state.
The occurrence of such a pure
  state can be understood from the analogy with the Rindler case, where
  boundary conditions corresponding to the presence of a mirror at the
  origin can be posed. Alternatively, one can consider a quantum state
  on the complete manifold \cite{FHM}. Tracing out the degrees of freedom
  of the left wedge would then lead to the well-known thermal
  density matrix in the right wedge. The derivation of the decoherence
  factor would proceed along the lines described in Section (\ref{deco}).

Another important problem is the possible occurrence of anomalies
which could spoil the whole semiclassical limit \cite{CJZ}.
This of course requires a full understanding of regularization
in quantum gravity. A proper regularization is also needed
for a rigorous evaluation of the correction terms to the
Schr\"odinger equation, which have been derived in Section~4.5.
 Maybe one can make use of the methods developed
in \cite{hor} and \cite{mans} where the attempt is made, in the
context of a strong coupling expansion, to define the
kinetic terms rigorously. One can also try to modify the expansion
scheme itself \cite{Kim}.

We also wish to comment briefly how our results are related to those
presented in \cite{klmo}.
There, a vacuum matter state was freely evolved on the
geometry of Figure~1 for a black hole of mass $M$ and then compared,
through an appropriate scalar product, to
the same evolution on a black hole of mass $M+\Delta M$.
It was claimed there that a value of $\Delta M / \lambda$  of the
order of $e^{- M /\lambda}$ would lead to a vanishing product, signalling
a breakdown of the semiclassical approximation
(if $\lambda$ is taken to be proportional to the inverse Planck length).
First we note that considering a scalar product is essentially what was
done in Section~5, although there we have not considered
superpositions with different masses.
But since the matter states describing Hawking radiation are identical for
two different black hole masses, we would reach the opposite
conclusion of an absence rather than an excess of decoherence
for superpositions of different masses.
It should be remembered, however,
 that the foliation chosen is quite different.
In our case, before reaching the horizon,
our spacelike slices extend to the left spatial infinity, while
 for times after
the shock wave, the matter evolution has support on the
coordinate system $v^\pm$ which only covers the region
outside the horizon (analogous to the Unruh effect).
In this way, the aspect of information for  half the line
being `lost' behind the
horizon is captured.  In \cite{klmo},
the matter states are compared on slices that cross the shock wave and
the horizon.
Given the different slicing studied, further work would be needed to clarify
how the two approaches are related.
{}From a practical point of view,
 we note the relative simplicity of our calculation
which makes use of the Schr\"odinger picture througout.

\ \ \ \ \ \ \ \\ \\ \\
\begin{centerline} {\bf ACKNOWLEDGMENTS} \end{centerline}
We would like  to thank Robert Myers for useful comments on the manuscript. 
\vspace{1cm}

\noindent

\vspace{1cm}


\vspace{1cm}

\newpage
\begin{thebibliography}{99}
\bibitem{stro} A. Strominger, ``Les Houches Lectures on
Black Holes", hep-th/9501071.
\bibitem{K} K. V. Kucha\v{r}, {\it Phys. Rev.} {\bf D50} (1994) 3961.
\bibitem{KT} H. A. Kastrup and
T. Thiemann, {\em Nucl. Phys.} B {\bf 425} (1994) 665.
\bibitem{rom} J. D. Romano, ``Spherical Symmetric Scalar Field
 Collapse: An example of the Spacetime Problem of Time",
 gr-qc/9501015.
\bibitem{kiefer1}{ C. Kiefer, in {\em Canonical Gravity: {}From classical
to quantum}, edited by J. Ehlers and H. Friedrich (Springer - Verlag,
Berlin, 1994).  }
\bibitem{KMS} C. Kiefer, R. M\"uller, and T. P. Singh,
 {\em Mod. Phys. Lett.} {\bf A9} (1994) 2661.
\bibitem{klmo}
E. Keski-Vakkuri, G. Lifschytz, S. D. Mathur, and M. E. Ortiz,
{\it Phys. Rev.} {\bf D51} (1995) 1764.
\bibitem{cghs}{ C.G. Callan, S.B. Giddings, J.A. Harvey, and A. Strominger,
 {\it Phys. Rev.} {\bf D45} (1992) R1005. }
\bibitem{kuns1} {  D. Louis-Martinez, J. Gegenberg,
 and G. Kunstatter, {\it Phys. Lett.} {\bf 321B}
(1994) 193. }
\bibitem{CJZ} D. Cangemi, R. Jackiw, and B. Zwiebach,
``Physical States in Matter-Coupled Dilaton Gravity",
to appear in {\em Annals of Physics}.
\bibitem{AI} S. P. de Alwis and D. A. MacIntire, {\it Phys. Rev.}
 {\bf D50} (1994) 5164.
\bibitem{kuns3}{ J. Gegenberg, G. Kunstatter, and D. Louis-Martinez,
{\it Phys. Rev.} {\bf D51} (1995) 1781.}
\bibitem{hori}{ T. Hori, {\it Prog. Theor. Phys.} {\bf 90} (1993) 743. }
\bibitem{dw}{B. DeWitt, {\it Phys. Rev.} {\bf 160} (1967) 1113.}
\bibitem{kuns2}{ D. Louis-Martinez and  G. Kunstatter,
{\it Phys. Rev.} {\bf D49} (1994) 5227. }
\bibitem{kiefer2} D. Giulini and C. Kiefer, {\em Class. Quantum Grav.}
 {\bf 12} (1995) 403.
\bibitem{GN} S. Giddings and W. Nelson, {\em Phys. Rev.} {\bf D46}
 (1992) 2486.
\bibitem{Ja} R. Jackiw, in {\em Field Theory and Particle Physics},
 edited by O. Eboli, M. Gomes, and A. Santano (World Scientific,
 Singapore, 1988).
\bibitem{GR} I. S. Gradshteyn and I. M. Ryzhik, {\em Table of Integrals,
 Series, and Products} (Academic Press, Orlando, 1980).
\bibitem{Haw}
S. W. Hawking, {\it Commun. Math. Phys.} {\bf 43} (1975) 199.
\bibitem{FHM} K. Freese, C. T. Hill, and M. Mueller,
  {\em Nucl. Phys.} {\bf B255} (1985) 693.
\bibitem{CM} M. Cadoni and S. Mignemi, {\it Phys. Lett.} {\bf 358B}
(1995) 217.
\bibitem{KS} C. Kiefer and T. P. Singh, {\em Phys. Rev.} {\bf D44}
 (1991) 1067.
\bibitem{Haj} P. H\'{a}j\'{\i}\v{c}ek,{\it Commun. Math. Phys.}
{\bf 150} (1992) 545.
\bibitem{Ki2} C. Kiefer, {\it Phys. Rev.} {\bf D46} (1992) 1658.
\bibitem{Ze} H. D. Zeh, in {\em Stochastic evolution of quantum states
 in open systems and measurement processes}, edited by L. Di\'{o}si
 and B. Luk\'{a}cs (World Scientific, Singapore, 1994).
\bibitem{hor} T. Horiguchi, K. Maeda, and M. Sakamoto,
 {\it Phys. Lett.} {\bf 344B} (1995) 105.
\bibitem{jgd} J.-G. Demers, {\it Mod. Phys. Lett.} {\bf A10}
(1995) 1745.
\bibitem{mans} P. Mansfield, {\it Nucl. Phys.} {\bf B148} (1994) 1130.
\bibitem{Kim} S. P. Kim, {\it Phys. Rev.} {\bf D52} (1995) 3382.
\end{thebibliography}
\end{document}